# Solving DC Power Flow Problems Using Quantum and Hybrid Algorithms

Fang Gao[a,b*], Guojian Wu[a,b], Suhang Guo[a,b], Wei Dai[a,b], Feng Shuang[a,b*]

[a] *Guangxi Key Laboratory of Intelligent Control and Maintenance of Power Equipment, Guangxi University, Nanning, 530004, China*

[b] *School of Electrical Engineering, Guangxi University, Nanning, 530004, China*

**Abstract**

Power flow calculation plays an important role in planning, operation, and control of the power system. The quantum HHL algorithm can achieve theoretical exponential speedup over classical algorithms on DC power flow calculation. Since the qubit resources in the Noisy Intermediate-scale Quantum (NISQ) era are limited, it is important to discuss the performance considering this limitation. The coefficient matrix of the linear systems of equations in DC power flow problems cannot be represented perfectly by finite binary number strings, which leads to imperfect phase estimation. This work is carried out under the assumption of imperfect phase estimation. The performance of the HHL algorithm is systematically investigated with different accuracy and redundant qubits. In order to further reduce the required qubit resources, a hybrid quantum-classical algorithm is proposed. By comparing errors of the HHL and hybrid algorithms in the DC power flow calculation of the IEEE 5-bus test system, it is found that the hybrid algorithm can achieve comparable precision with fewer qubits than HHL by increasing the number of phase estimation modules, which may make the hybrid algorithm a feasible route in the NISQ era.

*Keywords:* DC power flow, quantum algorithm, hybrid algorithm, qubit resources

∗ Corresponding author

*E-mail address:* fgao@gxu.edu.cn (F. Gao), gjwu@st.gxu.edu.cn (G. Wu), 1363054525@qq.com (S. Guo), weidai2019@163.com (W. Dai), fshuang@gxu.edu.cn (F. Shuang).

## 1. Introduction

With the development of quantum hardware, quantum computation is now becoming a promising computation paradigm. Algorithms such as HHL [1], Shor's factorization [2] and Grover [3] search implemented on quantum hardware make use of quantum characteristics (i.e. superposition and entanglement) to achieve quantum advantages. A large-scale error corrected quantum computer can solve problems that even the largest classical supercomputers can not. However, we are now in the so-called noisy intermediate-scale quantum (NISQ) era in which quantum hardware still has limitations in the qubit resources (i.e. the depth of quantum circuits and the number of qubits). Therefore, another trend is to develop quantum-classical hybrid algorithms [4–17] to lower the requirement of qubit resources by combining quantum computation and classical computation. Quantum-inspired algorithms [18–29] run on classical computers and introduce some quantum features into classical algorithms. These two types of algorithms can also bring potential performance improvement. Development and application of the aforementioned three types of quantum-related algorithms have attracted much attention, and are finding applications in many scenarios such as power systems [30–36].

Power flow calculation is the most basic and important calculation in power systems [37–39]. It is to obtain the bus voltage, branch power and network loss during the steady-state operation of the power system, knowing the connection mode, parameters and operation conditions of the power grid. Nowadays, the permeability of new energy in the power grid is increasing due to environmental and energy considerations. New energy sources such as wind and solar energies are random and intermittent, making the mathematical model of the large power grid extremely complex. The real-time uncertainty analysis of the grid is posing a great challenge for the existing algorithms since tremendous number of repetitive power flow calculations are involved. In particular, the complexity of $N$-1 security check, which has been widely used in many scenarios such as security-constrained economic dispatching, security-constrained unit commitment and power market clearing, is also increasing exponentially. For example, if $N$-1 security check is carried out without new energy sources, $N$ times power flow calculations are needed; when there are $M$ new energy power plants in the power grid with each plant containing $x$ scenarios, $(N-1)M^x$ times calculations are needed.

In some specific scenarios, the AC power flow model is usually simplified to the DC power flow model [40–43]. Although the solution accuracy of the DC model is not as good as that of the AC model, it is very advantageous in scenarios (i.e. overload check calculation) where high accuracy is not required and the reactive power flow is very small compared with the active power flow. This paper focuses on the application of quantum and hybrid algorithms in solving DC power flow equations.

DC power flow equations are linear equations. The solution of linear equations is one basic problem in linear algebra and many classical algorithms have been developed. Compared with classical algorithms, the HHL algorithm proposed by Harrow et al in 2009 can achieve exponential acceleration theoretically in solving the linear system of equations under certain conditions [1],[44]. Eskandarpour et al also proved that the HHL algorithm is superior to classical algorithms in solving DC power flow problems [45]. Feng et al proposed an enhanced HHL algorithm to solve the fast decoupled quantum power flow [46].

The original HHL algorithm can be improved in the following three aspects[47],[48]: (1) A more efficient initial state preparation scheme. When using HHL algorithm to solve the system of linear equations $B\theta = P$ ( $B$ is the coefficient matrix and $P$ is a known column vector), it is necessary to map $P$ to the quantum register, that is, to prepare an initial state whose probability amplitude distribution corresponds to $P$; (2) Schemes to reduce the degree of restraint by ill-conditioned matrix. When the matrix $B$ is ill-conditioned, the accuracy of the original HHL algorithm will be greatly affected and may output an incorrect solution. (3) Schemes to reduce the requirement of quantum resources. The original HHL algorithm guarantees the correctness and precision of the solution by allocating more quantum resources (i.e. more qubits), and insufficient available quantum resources may lead to unsatisfactory performance of the HHL algorithm.

Different schemes are proposed to improve the HHL algorithm concerning about the above three aspects. Long et al. proposed a scheme for preparing arbitrary quantum states [49]. Harrow delivered an improved phase estimation scheme [1], and a filtering function is introduced to filter out the influence caused by the ill part of the matrix in order to increase the reliability of the algorithm. Lee et al proposed a hybrid HHL algorithm [50], which reduced the depth of the quantum circuit by processing some information with a classical computer. This hybrid scheme improved the performance of the HHL algorithm since the accumulated error will be reduced with shallower quantum circuits. However, the number of qubits allocated in Lee's hybrid HHL algorithm was the same as the original HHL algorithm and the number of required qubits was not reduced.

When the HHL algorithm under imperfect phase estimation is applied to DC power flow calculation, huge qubit resources are needed for accuracy consideration, which hinders its application in the NISQ era [51]. Zhou et al devised an iteration-based HHL revision (i.e. quantum electromagnetic transients program) for mitigating temporal errors to achieve high accuracy under limited qubit resources [52]. When the eigenvalues of the coefficient matrix of the linear systems of equations can be represented perfectly by finite binary number strings (i.e. perfect phase estimation), HIPEA [53] was designed in our previous work based on the iterative phase estimation algorithm to lower the number of required qubits. However, the eigenvalues of the admittance matrix $B$ in the DC power flow problems can hardly be perfectly represented by a finite-bit binary number, so the corresponding phase estimation is imperfect leading to a failure probability [1]. This paper focuses on the case of imperfect phase estimation and develops a new hybrid quantum-classical algorithm to solve the DC power flow equation. The proposed algorithm can reduce the demand for qubit resources without sacrificing accuracy. The IEEE 5-bus test system is taken as an example for error analysis of the HHL and hybrid algorithms under different resource consumptions.

The remaining part of this paper is organized as follows. In Sec. 2, the HHL and hybrid algorithms under imperfect phase estimation are presented. Results of the example system and the corresponding discussions are given in Sec. 3. Sec. 4 is the conclusion.

## 2. Quantum and Hybrid Algorithms for Solving DC Power Flow Problems

The DC power flow equation is $P = B\theta$, where $P$ is the nodal active power injection, $B$ is the admittance matrix, and $\theta$ is the phase angle to be solved. In both quantum and hybrid algorithms, $P$ needs to be normalized and mapped to the quantum state $|C_P \cdot P\rangle$, where $C_P$ is the normalized constant, and the linear system of equations to be solved becomes $C_P P = C_P B\theta$. For the quantum HHL algorithm, the solution is stored in the quantum register and recorded as $|C_\theta \cdot \theta\rangle$, where $C_\theta$ is the normalized constant of $\theta$. Then the system of equations to be solved becomes $|C_P \cdot P\rangle = C_P C_\theta^{-1} B |C_\theta \cdot \theta\rangle$.

### 2.1 HHL under Imperfect Phase Estimation

Because the eigenvalues of the admittance matrix $B$ in the DC power flow problems can hardly be perfectly represented by a finite-bit binary number, the corresponding phase estimations are imperfect resulting in a failure probability [1]. Therefore, eigenvalues with error will be obtained. The quantum circuit of the HHL algorithm under imperfect phase estimation is shown in Fig. 1, which includes the top register, the medium register and the bottom register. The $m$ qubits in the medium register are divided into $n_{accur}$ accuracy qubits and $n_{redun}$ redundant qubits, which are, respectively, used to determine the accuracy and improve the success rate of phase estimations. The failure rate bound $\epsilon$ of phase estimations can be reduced by increasing $n_{redun}$, and the relationship is $\epsilon \leq (2 \cdot (2^{n_{redun}} - 2))^{-1}$ [54].

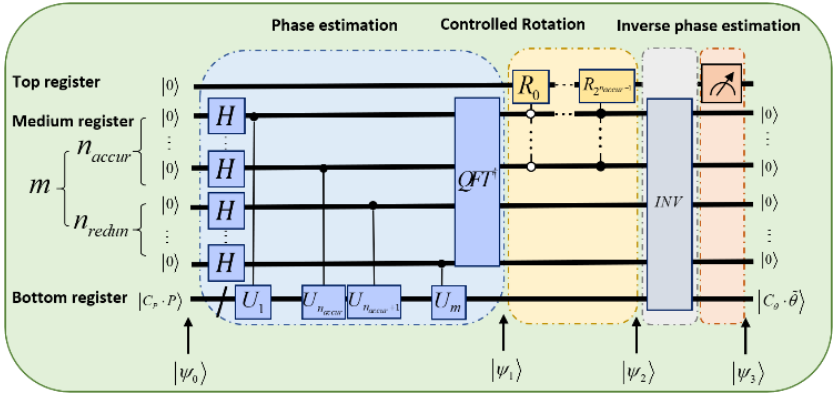

**Fig. 1.** Quantum circuit of HHL under imperfect phase estimation. The $k$-th unitary and rotation gates are, respectively, $U_k = e^{2\pi i B \cdot 2^{m-k}}$ and $R_k = R_y(\arcsin(2k))$.

The quantum circuit consists of the following three modules:

· *Phase Estimation Module:* During the phase estimation, the quantum state $|C_P \cdot P\rangle$ of the qubits in the bottom register is decomposed in the eigenspace of the $N$-dimensional matrix $B$ as $|C_P \cdot P\rangle = \sum_{j=1}^{N} C_P p_j |u_j\rangle$, where $|u_j\rangle$ is the unit eigenvector of the admittance matrix $B$ and $C_P p_j$ is the projection coefficient of $|C_P \cdot P\rangle$ on eigenvector $|u_j\rangle$, which means that $C_P p_j = \langle u_j | C_P \cdot P\rangle$. The high $m$ bits of the eigenvalue $\lambda_j$ in binary representation are encoded into the basis states of the first $m$ qubits in the medium register, which is marked as $|\lambda_j^{[1:m]}\rangle$. After the phase estimation, the system will evolve from the initial state

$$|\psi_0\rangle = |0\rangle|0\rangle^{\otimes m}|C_P \cdot P\rangle \tag{1}$$

to

$$|\psi_1\rangle = C_P |0\rangle \sum_{j=1}^{N} p_j |\lambda_j^{[1:m]}\rangle |u_j\rangle. \tag{2}$$

· *Controlled Rotation Module:* The controlled rotation module transmits the eigenvalue information encoded in the qubits in the medium register to the amplitude of the state of the qubit in the top register in reciprocal form through a series of controlled rotation gates. It is worth noted that the HHL algorithm under perfect phase estimation takes all the qubits in the medium register as the control qubits, while under imperfect phase estimation it takes $n_{accur}$ accuracy qubits as the control qubits to store the high $n_{accur}$ bits of the eigenvalue $\lambda_j$ in binary representation, noted as $|\lambda_j^{[1:n_{accur}]}\rangle$. After controlled rotation, the state of the system evolves to

$$|\psi_2\rangle = C_P \sum_{j=1}^{N} p_j \left( \sqrt{1 - \left( \frac{C_\lambda}{\lambda_j^{[1:n_{accur}]}} \right)^2} \; |0\rangle \; + \frac{C_\lambda}{\lambda_j^{[1:n_{accur}]}} |1\rangle \right) |\lambda_j^{[1:m]}\rangle |u_j\rangle, \tag{3}$$

where $C_\lambda$ is the normalized constant and satisfy $(C_P C_\lambda)^2 \sum_{j=1}^{N} (p_j / \lambda_j^{[1:n_{accur}]})^2 = 1$.

· *Inverse Phase Estimation Module:* Through the inverse phase estimation, the qubits in the medium register are unentangled with the other qubits. Then by the post-selection [55], the amplitude $C_P C_\lambda p_j / \lambda_j^{[1:n_{accur}]}$ is passed to $|u_j\rangle$ when the qubit in the top register is measured as $|1\rangle$. Ignoring the qubit in the top register, the state of the system is

$$|\psi_3\rangle = C_P C_\lambda \sum_{j=1}^{N} \frac{p_j |0\rangle^{\otimes m} |u_j\rangle}{\lambda_j^{[1:n_{accur}]}}, \tag{4}$$

and the quantum state stored in the bottom register is just the normalized solution

$$|C_P \cdot \tilde{\theta}\rangle = C_P C_\lambda \Sigma_{j=1}^N \frac{p_j |u_j\rangle}{\lambda_j^{[1:n_{accur}]}} .$$ (5)

There are two factors in the above process leading to deviation of the calculated $|C_\theta \cdot \tilde{\theta}\rangle$ from the true $|C_\theta \cdot \theta\rangle$: the failure rate of the phase estimation determined by $n_{redun}$ and the eigenvalue truncation error determined by $n_{accur}$. Without losing generality, it is assumed here $0 < \lambda_j < 1$, whose binary form is expressed as $\lambda_j = 0.\lambda_{j1}\lambda_{j2}\cdots$. Then $n_{accur}$ determines that the theoretical precision of $\lambda_j^{[1:n_{accur}]}$ is $2^{-n_{accur}}$, which means $\lambda_j^{[1:n_{accur}]} = \lfloor \lambda_j 2^{n_{accur}} \rfloor 2^{-n_{accur}}$, with "$\lfloor x \rfloor$" denoting the largest integer smaller than or equal to $x$. The approximate solution by HHL is:

$$\begin{aligned}|C_\theta \cdot \tilde{\theta}\rangle_{theory} &= C_P C_\lambda \Sigma_{j=1}^N \frac{p_j |u_j\rangle}{\lambda_j^{[1:n_{accur}]}} \\ &= C_P C_\lambda \Sigma_{j=1}^N \frac{p_j |u_j\rangle}{\lfloor \lambda_j 2^{n_{accur}} \rfloor \cdot 2^{-n_{accur}}}\end{aligned} .$$ (6)

Theoretically the corresponding relative error is

$$n_\varepsilon^{theory} = \frac{\left\| |C_\theta \cdot \tilde{\theta}\rangle_{theory} - |C_\theta \cdot \theta\rangle \right\|}{\left\| |C_\theta \cdot \theta\rangle \right\|} ,$$ (7)

while the relative error in our numerical experiment is

$$n_\varepsilon^{exp} = \frac{\left\| |C_\theta \cdot \tilde{\theta}\rangle - |C_\theta \cdot \theta\rangle \right\|}{\left\| |C_\theta \cdot \theta\rangle \right\|}$$ (8)

## 2.2 Hybrid Single Phase Estimation Algorithm

As shown in Fig. 2, we can also solve the DC power flow equation by a hybrid single phase estimation algorithm, namely HSPEA. The quantum circuit contains a top register and a bottom register, and the qubits in the top register are also divided into accuracy and redundant qubits. The accuracy qubits in the top register and the qubits in the bottom register are measured after the phase estimation to extract the information needed to solve the DC power flow equation. A Lemma, which is the theoretical basis of Step 1 of HSPEA, is first introduced and then proved in two different ways in the following.

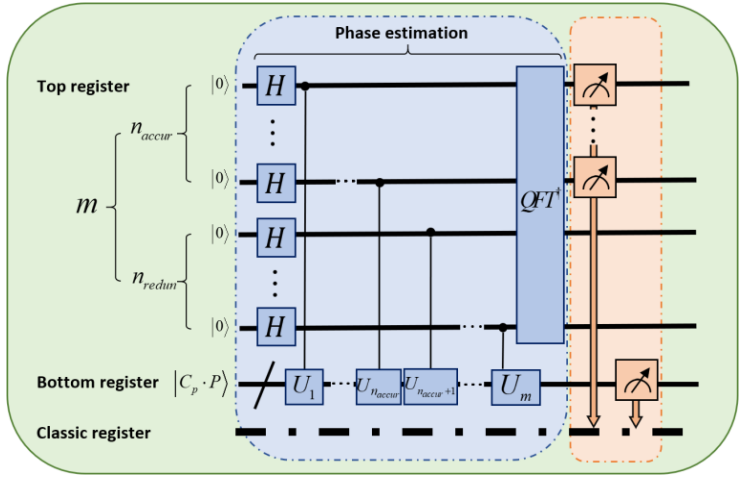

**Fig. 2.** Quantum circuits of the hybrid algorithm HSPEA

*Lemma:* After the phase estimation, the accuracy qubits in the top register is measured as $|\lambda_j^{[1:n_{accur}]}\rangle$ with the probability of $C_P^2 p_j^2$.

*Proof 1:* In HSPEA, when the redundant qubits are not considered, we denote the system state after phase estimation as

$$|\psi_1^S\rangle = \Sigma_{j=1}^N C_P p_j |\lambda_j^{[1:n_{accur}]}\rangle |u_j\rangle . \tag{9}$$

After the measurement, the accuracy qubits in the top register collapse to $|\lambda_{j'}^{[1:n_{accur}]}\rangle$ with the theoretical probability amplitude as

$$\begin{aligned}
&\Sigma_{i=1}^N \langle u_i|\langle \lambda_{j'}^{[1:n_{accur}]}|\psi_1^S\rangle \\
&= \Sigma_{j=1}^N (\Sigma_{i=1}^N \langle u_i|\langle \lambda_{j'}^{[1:n_{accur}]}|) C_P p_j |\lambda_j^{[1:n_{accur}]}\rangle |u_j\rangle
\end{aligned} . \tag{10}$$

According to post-selection, we know that

$$\langle u_i|\langle \lambda_{j'}^{[1:n_{accur}]}|\psi_1^S\rangle = \begin{cases} 0, & i \neq j' \\ \langle u_{j'}|\langle \lambda_{j'}^{[1:n_{accur}]}|\psi_1^S\rangle, & i = j' \end{cases} . \tag{11}$$

Therefore, Eq. (10) can be simplified as

$$\begin{aligned}
&\Sigma_{i=1}^N \langle u_i|\langle \lambda_{j'}^{[1:n_{accur}]}|\psi_1^S\rangle \\
&= \langle u_{j'}|\langle \lambda_{j'}^{[1:n_{accur}]}|\psi_1^S\rangle \\
&= \Sigma_{j=1}^N C_P p_j \langle u_{j'}|\langle \lambda_{j'}^{[1:n_{accur}]}|\lambda_j^{[1:n_{accur}]}\rangle |u_j\rangle
\end{aligned} . \tag{12}$$

The eigenvalues are encoded in basis states, so we have

$$\langle \lambda_{j'}^{[1:n_{accur}]}|\lambda_j^{[1:n_{accur}]}\rangle = \begin{cases} 0, & j \neq j' \\ 1, & j = j' \end{cases} , \tag{13}$$

leading to the following probability amplitude

$$\begin{aligned}
&\Sigma_{i=1}^N \langle u_i|\langle \lambda_{j'}^{[1:n_{accur}]}|\psi_1^S\rangle \\
&= C_P p_{j'} \langle u_{j'}|\langle \lambda_{j'}^{[1:n_{accur}]}|\lambda_{j'}^{[1:n_{accur}]}\rangle |u_{j'}\rangle \\
&= C_P p_{j'}
\end{aligned} . \tag{14}$$

So we can draw the conclusion that the probability of the measurement result $|\lambda_{j'}^{[1:n_{accur}]}\rangle$ is $C_P^2 p_{j'}^2$ and complete the proof.

*Proof 2:* Besides the above proof, we also give a proof without using post-selection theory.

According to Eq. (13), Eq. (10) can be transformed as follows:

$$\begin{aligned}
&\Sigma_{i=1}^N \langle u_i|\langle \lambda_{j'}^{[1:n_{accur}]}|\psi_1^S\rangle \\
&= \Sigma_{j=1}^{j'-1} (C_P p_j \Sigma_{i=1}^N (\langle u_i|\lambda_{j'}^{[1:n_{accur}]}|\lambda_j^{[1:n_{accur}]}\rangle |u_j\rangle)) \\
&\quad + C_P p_{j'} \Sigma_{i=1}^N \langle u_i|\lambda_{j'}^{[1:n_{accur}]}|\lambda_{j'}^{[1:n_{accur}]}\rangle |u_{j'}\rangle \\
&\quad + \Sigma_{j=1}^{j'+1} (C_P p_j \Sigma_{i=1}^N (\langle u_i|\lambda_{j'}^{[1:n_{accur}]}|\lambda_j^{[1:n_{accur}]}\rangle |u_j\rangle)) \\
&= C_P p_{j'} \Sigma_{i=1}^N \langle u_i|\lambda_{j'}^{[1:n_{accur}]}|\lambda_{j'}^{[1:n_{accur}]}\rangle |u_{j'}\rangle \\
&= C_P p_{j'} \Sigma_{i=1}^N \langle u_i|u_{j'}\rangle
\end{aligned} . \tag{15}$$

Due to the orthogonality between eigenvectors

$$\langle u_i|u_{j'}\rangle = \begin{cases} 0, & i \neq j' \\ 1, & i = j' \end{cases} , \tag{16}$$

the probability amplitude is

$$\begin{aligned}
&\Sigma_{i=1}^N \langle u_i|\langle \lambda_{j'}^{[1:n_{accur}]}|\psi_1^S\rangle \\
&= C_P p_{j'} (\Sigma_{i=1}^{j'-1} \langle u_i|u_{j'}\rangle + \langle u_{j'}|u_{j'}\rangle + \Sigma_{i=1}^{j'+1} \langle u_i|u_{j'}\rangle) \\
&= C_P p_{j'}
\end{aligned} . \tag{17}$$

Obviously, Eq. (14) and Eq. (17) give the same result.

So we can draw the conclusion that the probability of the measurement result $|\lambda_j^{[1:n_{acur}]}\rangle$ is $C_P^2 p_j^2$ and complete the proof.

As shown in Algorithm 1, the main steps of HSPEA are as follows:

*Step 1:* According to the *Lemma*, the value of $\lambda_j^{[1:n_{acur}]}$ and its corresponding $|C_P p_j|$ will be obtained after the phase estimation.

*Step 2:* According to the post-selection mentioned above, when the accuracy qubits in the top register are measured as $|\lambda_j^{[1:n_{acur}]}\rangle$, the qubits in the bottom register will collapse to the eigenstate $|u_j\rangle$, with $u_{jq}$ as its $q$-th element. Since $u_{jq}^2$ can be obtained as the probability of the corresponding product state of the qubits in the bottom register (i.e. $u_{j1}^2$ is the probability of $|0\rangle|0\rangle\cdots|0\rangle$, $u_{j2}^2$ is the probability of $|0\rangle|0\rangle\cdots|1\rangle$), $|u_{jq}|$ can be obtained.

*Step 3:* The positive and negative sign information of each element $p_j u_{jq}$ is calibrated by $|C_P P\rangle = C_P \Sigma_{j=1}^N p_j |u_j\rangle$ with the following procedures:

$|C_P P\rangle = C_P \Sigma_{j=1}^N p_j |u_j\rangle$ can be expanded into the form of

$$
\begin{aligned}
C_P P_1 &= (-1)^{n_{11}} \times C_P \left|p_1 u_{11}\right| + (-1)^{n_{21}} \times C_P \left|p_2 u_{21}\right| + \cdots + (-1)^{n_{N1}} \times C_P \left|p_N u_{N1}\right| \\
C_P P_2 &= (-1)^{n_{12}} \times C_P \left|p_1 u_{12}\right| + (-1)^{n_{22}} \times C_P \left|p_2 u_{22}\right| + \cdots + (-1)^{n_{N2}} \times C_P \left|p_N u_{N2}\right| \\
&\vdots \\
C_P P_N &= (-1)^{n_{1N}} \times C_P \left|p_1 u_{1N}\right| + (-1)^{n_{2N}} \times C_P \left|p_2 u_{2N}\right| + \cdots + (-1)^{n_{NN}} \times C_P \left|p_N u_{NN}\right|
\end{aligned}
\tag{18}
$$

where $P_q$ is the $q$-th element of $P$, and $n_{jq}$ is an integer of 0 or 1 indicating the positive or negative sign of $n_{jq}$. In Eq. (18), values of $C_P P_q$ and $C_P |p_j u_{jq}|$ for all $q, j \in N$ are known, which means that the information of signs can be determined by traversing all possible (i.e. positive or negative) combinations of $n_{jq}$. So $p_j u_{jq}$ and thus $p_j u_j$ (i.e. $u_j$ is a vector with elements $u_{jq}$) can be obtained.

*Step 4:* The solution of the DC power flow equation is calculated in the classical register as

$$
\tilde{\theta} = \Sigma_{j=1}^N \frac{p_j u_j}{\lambda_j^{[1:n_{acur}]}} .
\tag{19}
$$

Compared with HHL, HSPEA's circuit depth is reduced.

---

**Algorithm 1:** Hybrid Single Phase Estimation Algorithm

**Input:** A coefficient matrix $B$ and a known column vector $P$

**Output:** A solution vector $\theta$

Initialize the quantum registers to $|0\rangle^{\otimes m}|C_P \cdot P\rangle$

1. Use $B$ to generate the unitary matrices ($U_i, i = 1, \ldots, m$) in phase estimation;
2. Perform phase estimation;
3. Measure the accuracy qubits in the top register and the qubits in the bottom register;
4. Count the measurement results of the top register to obtain $|\lambda_j^{[1:n_{acur}]}\rangle$ and $|C_P p_j|$;
5. Count the measurement results of the top and bottom registers to obtain $|u_{jq}|$;
6. Calibrate the positive and negative sign information of each element $p_j u_{jq}$ according to equation $|C_P P\rangle = C_P \Sigma_{j=1}^N p_j |u_j\rangle$, and then obtain $p_j u_j$;
7. Calculate the solution vector $\theta$ according to $p_j u_j$ and $\lambda_j^{[1:n_{acur}]}$ obtained from the above steps;
8. Output $\theta$.

## 2.3 Hybrid Multiple Phase Estimation Algorithm

Inspired by the idea of multiple iterative modules of HIPEA [53], a hybrid multiple phase estimation algorithm (HMPEA) is developed based on HSPEA. Fig. 3 shows the quantum circuit of HMPEA, where $m_{prec}$ represents the estimation precision of the eigenvalues (i.e. $m_{prec}$ bits starting from the highest bit in the binary representation are estimated). Each phase estimation module estimates $n_{accur}$ bits of the eigenvalue, and the eigenvalue can be obtained by $m_{prec} / n_{accur}$ phase estimation modules assuming that $m_{prec}$ is divisible by $n_{accur}$.

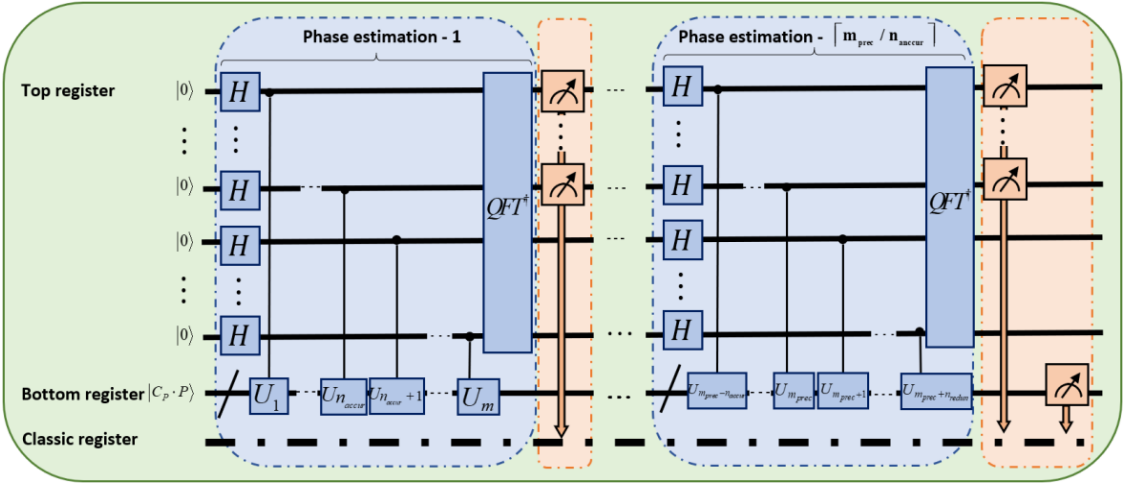

**Fig. 3.** Quantum circuits of the hybrid algorithm HMPEA.

Generally, the number of phase estimation modules is $\lceil m_{prec} / n_{accur} \rceil$, where "$\lceil x \rceil$" means the smallest integer greater than or equal to $x$. The overall steps of HMPEA are almost the same as HSPEA, and the difference mainly lies in the scheme of extracting eigenvalues and the corresponding coefficients. To show the features of HMPEA more clearly, we represent the $N$ different eigenvalues (i.e. $\lambda_j = 0.\lambda_{j1}\lambda_{j2}\cdots\lambda_{jm_{prec}}\lambda_{jm_{prec}+1}\cdots$) of the coefficient matrix $B$ in the form of a binary tree in Table 1. Here the subscript of $\lambda$ represents the index number of the eigenvalue while the superscript represents the binary digit range of $\lambda$. For example, $\lambda_{\{1,2,\cdots,N\}}^{\{1:n_{accur}\}}$ means the high $n_{accur}$ bits of all the eigenvalues. If values from $(d_x-1)n_{accur}+1$ to $d_x n_{accur}$ bits of any two eigenvalues differ, the $x$-th divergence appears in the $d_x$-th phase estimation module [53]. Two groups of eigenvalues can be distinguished at each divergence. Table 1 shows that before the $d_1$-th phase estimation module, the measurement results of the accuracy qubits in the top register are the same [56]. The first divergence appears in the $d_1$-th phase estimation module (i.e. two different measurement results are obtained), and the eigenvalues are divided into two groups ($\lambda_{\{1,2,\cdots,n_{d1}\}}$ and $\lambda_{\{n_{d1}+1,n_{d1}+2,\cdots,N\}}$) with $\lambda_{n_{d1}}$ as the boundary. Finally, after the $(N-1)$-th divergence (i.e. the $d_{N-1}$-th phase estimation module), all the $N$ eigenvalues can be differentiated.

As shown in Table 1, each phase estimation module extracts $n_{accur}$ bits of the eigenvalue, and $m_{prec}$ bits of the eigenvalues and the corresponding joint probabilities (i.e. $\lambda_j^{\{1:m_{prec}\}}$ and $|C_p p_j|$) are obtained after the last phase estimation module.

**Table 1**

A binary tree representation of $N$ different eigenvalues (i.e. $\lambda_j$) with their corresponding joint probabilities. The joint probability in the $d_x$-th phase estimation can be obtained by measuring accuracy qubits in phase estimation modules from 1 to $d_x$.

| Phase Estimation - index | | $\lambda_1$ | $\lambda_2$ | $\cdots$ | $\lambda_{N-1}$ | $\lambda_N$ |
|---|---|---|---|---|---|---|
| Phase Estimation - 1 | Eigenvalues | $\lambda_{\{1,2,\cdots,N\}}^{[1:n_{accur}]}$ | | | | |
| | Probability | $\Sigma_{j=1}^N C_P^2 p_j^2 = 1$ | | | | |
| Phase Estimation - 2 | Eigenvalues | $\lambda_{\{1,2,\cdots,N\}}^{[1:2n_{accur}]}$ | | | | |
| | Joint probability | $\Sigma_{j=1}^N C_P^2 p_j^2 = 1$ | | | | |
| $\vdots$ | | | | | | |
| Phase Estimation - $d_1$ | Eigenvalues | $\lambda_{\{1,2,\cdots,n_{d1}\}}^{[1:d_1 n_{accur}]}$ | | | $\lambda_{\{n_{d1}+1,n_{d1}+2,\cdots,N\}}^{[1:d_1 n_{accur}]}$ | |
| | Joint probability | $\Sigma_{j=1}^{n_{d1}} C_P^2 p_j^2$ | | | $\Sigma_{j=n_{d1}+1}^N C_P^2 p_j^2$ | |
| $\vdots$ | | | | | | |
| Phase Estimation - $d_{N-1}$ | Eigenvalues | $\lambda_1^{[1:d_{N-1} n_{accur}]}$ | $\lambda_2^{[1:d_{N-1} n_{accur}]}$ | $\cdots$ | $\lambda_{N-1}^{[1:d_{N-1} n_{accur}]}$ | $\lambda_N^{[1:d_{N-1} n_{accur}]}$ |
| | Joint probability | $C_P^2 p_1^2$ | $C_P^2 p_2^2$ | | $C_P^2 p_{N-1}^2$ | $C_P^2 p_N^2$ |
| $\vdots$ | | | | | | |
| Phase Estimation - $\lceil m_{prec}/n_{accur}\rceil$ | Eigenvalues | $\lambda_1^{[1:m_{prec}]}$ | $\lambda_2^{[1:m_{prec}]}$ | $\cdots$ | $\lambda_{N-1}^{[1:m_{prec}]}$ | $\lambda_N^{[1:m_{prec}]}$ |
| | Joint probability | $C_P^2 p_1^2$ | $C_P^2 p_2^2$ | | $C_P^2 p_{N-1}^2$ | $C_P^2 p_N^2$ |

---

**Algorithm 2:** Hybrid Multiple Phase Estimation Algorithm

**Input:** A coefficient matrix $B$ and a known column vector $P$

**Output:** A solution vector $\theta$

Initialize the quantum registers to $|0\rangle^{\otimes m}|C_P \cdot P\rangle$

1. Use $B$ to generate the unitary matrices ($U_i, i=1,\ldots,m$) in phase estimation;
2. **for** $i=1:\lceil m_{prec}/n_{accur}\rceil$ **do**
3.     Perform the $i$-th phase estimation;
4.     Measure the accuracy qubits in the top register;
5.     **if** $i=\lceil m_{prec}/n_{accur}\rceil$ **do**
6.         Measure the qubits in the bottom register;
7.     **end if**
8.     Initialize the top and bottom registers;
9. **end for**
10. Count the measurement results of the top register to obtain $|\lambda_j^{[1:n_{accur}]}\rangle$ and $|C_P p_j|$;
11. Count the measurement results of the top and bottom registers to obtain $|u_{j q}|$;
12. Calibrate the positive and negative sign information of each element $p_j u_{j q}$ according to equation $|C_P P\rangle = C_P \Sigma_{j=1}^N p_j |u_j\rangle$, and then obtain $p_j u_j$;
13. Calculate the solution vector $\theta$ according to $p_j u_j$ and $\lambda_j^{[1:n_{accur}]}$ obtained from the above steps;
14. Output $\theta$.

As shown in Algorithm 2, the main steps of HMPEA are as follows:

*Step 1:* After executing multiple phase estimation modules, each eigenvalue together with its joint probability (i.e. $\lambda_j^{[1:m_{prec}]}$ and $\left|C_P p_j\right|$ ) is obtained by measurement.

*Step 2:* The qubits in the bottom register are only measured after the last module to obtain $\left|u_{jq}\right|$.

*Step 3:* The same scheme in Step 3 of HSPEA is used to calibrate the positive and negative sign information of each element $p_j u_{jq}$.

*Step 4:* The solution is obtained according to Eq. (19).

The success rate of phase estimation in HMPEA is also a source of error. Each phase estimation module estimates the $n_{accur}$ bits of the eigenvalue with the lowest bound of success rate as

$$P_{success}^S = 1 - \frac{1}{2 \cdot (2^{n_{redun}} - 2)}, \tag{20}$$

where the superscript *S* denotes a single module. When the statistical measurement error of the probability is not considered, the correct eigenvalue is obtained by $\lceil m_{prec} / n_{accur} \rceil$ phase estimation modules with the lowest bound of success rate as

$$P_{success}^M = \left(1 - \frac{1}{2 \cdot (2^{n_{redun}} - 2)}\right)^{\left\lceil \frac{m_{prec}}{n_{accur}} \right\rceil}, \tag{21}$$

where the superscript *M* denotes multiple modules. Fig. 4 shows how $P_{success}^M$ changes with $n_{accur}$ and $n_{redun}$ in the case of $m_{prec} = 9$.

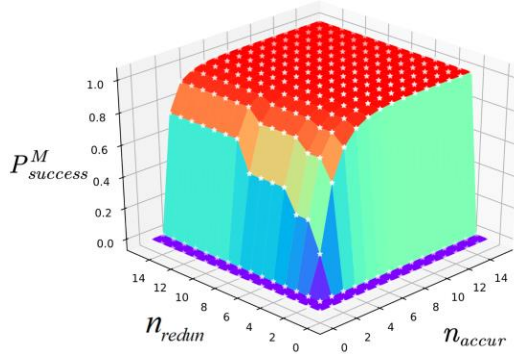

**Fig. 4.** The curved surface of $P_{success}^M$ in the case of $m_{prec} = 9$. The white stars denote results with different $n_{accur}$ and $n_{redun}$.

The error of the solution is related to the failure possibility of phase estimation and the required solution accuracy. When both $m_{prec}$ and $m$ (i.e. $n_{accur} + n_{redun}$) are 9, the lowest bound of the overall success rate of phase estimation is

$$P_{success}^M = \left(1 - \frac{1}{2 \cdot (2^{9-n_{accur}} - 2)}\right)^{\left\lceil \frac{9}{n_{accur}} \right\rceil}. \tag{22}$$

The curves of $P_{success}^M$ and $n_{module} = \lceil m_{prec} / n_{accur} \rceil$ versus $n_{accur}$ are shown in Fig. 5. $P_{success}^M$ decreases monotonously with the increase of $n_{accur}$ (the orange line), but a smaller $n_{accur}$ leads to more phase estimation modules $n_{module}$ (the blue line), which means a higher time complexity. So reasonable allocation of available qubit resources will effectively improve the performance (both in accuracy and time complexity) of HMPEA.

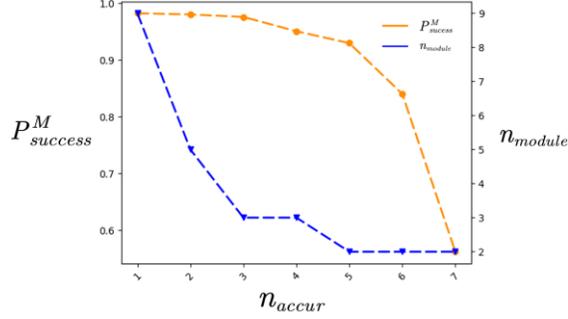

**Fig. 5.** Curves of $P_{success}^{M}$ and $n_{module}$ versus $n_{accur.}$ in HMPEA for $m = 9$ and $m_{prec} = 9$.

When only the truncation error is considered, the theoretical solution is

$$
\begin{aligned}
\tilde{\theta}_{theory} &= \sum_{j=1}^{N} \frac{p_j |u_j\rangle}{\lambda_j^{[1:m_{prec}]}} \\
&= \sum_{j=1}^{N} \frac{p_j |u_j\rangle}{\lfloor \lambda_j 2^{m_{prec}} \rfloor \cdot 2^{-m_{prec}}}
\end{aligned}
\tag{23}
$$

The corresponding theoretical relative error is

$$
n_{c}^{throry} = \frac{\left\| \tilde{\theta}_{theory} - \theta \right\|}{\left\| \theta \right\|},
\tag{24}
$$

while the relative error in our numerical experiment is

$$
n_{c}^{\exp} = \frac{\left\| \tilde{\theta} - \theta \right\|}{\left\| \theta \right\|}.
\tag{25}
$$

## 3. Results and Discussions

In the following, the IEEE 5-bus test system is taken as an example for illustration, and the model is shown in Fig. 6.

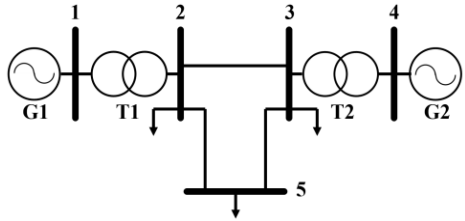

**Fig. 6.** IEEE 5-bus test system

The DC power flow equations corresponding to the model are as follows:

$$
\begin{bmatrix}
224.7319 & -35.5872 & 0 & -156.25 \\
-35.5872 & 128.1798 & -92.5926 & 0 \\
0 & -92.5926 & 126.2626 & 0 \\
-156.25 & 0 & 0 & 189.92
\end{bmatrix}
\theta =
\begin{bmatrix}
-0.1113 \\
-0.2623 \\
0.3169 \\
0.9046
\end{bmatrix}.
\tag{26}
$$

The solution obtained by the classical algorithm is

$$\theta = \begin{bmatrix} 0.0082 & 0.0043 & 0.0057 & 0.0115 \end{bmatrix}^T . \tag{27}$$

The corresponding normalized solution is

$$|C_\theta \theta\rangle = \begin{bmatrix} 0.5173 & 0.2740 & 0.3595 & 0.7267 \end{bmatrix}^T . \tag{28}$$

The quantum simulator PyQPanda [57] is employed in our numerical experiments, and the qubits needed for different precisions (expressed in the binary representation) with the HHL and HMPEA algorithms are shown in Fig. 7.

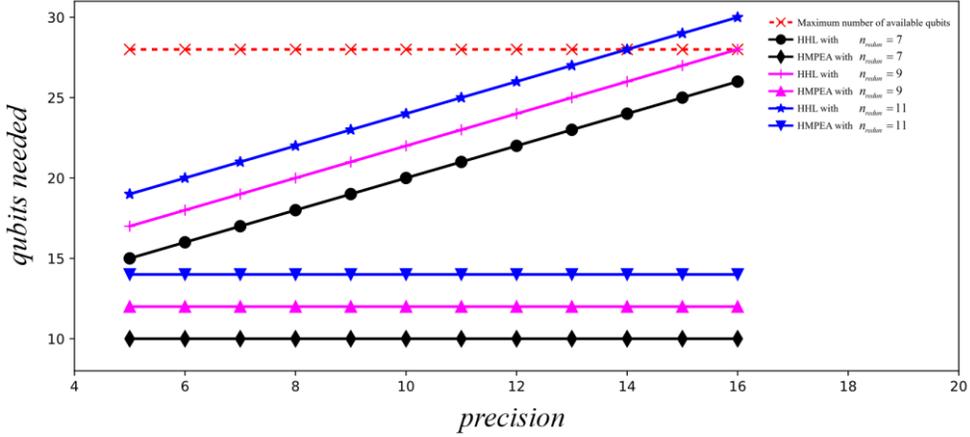

**Fig. 7.** Qubits needed for different precisions

In HHL, the precision is $2^{-n_{accur}}$, and the number of qubits needed is $n_{top} + n_{accur} + n_{redun} + n_{bottom}$, where $n_{top}$ and $n_{bottom}$ are, respectively, the qubits allocated for the top register and bottom register. In our experiments, $n_{top}$ and $n_{bottom}$ are two fixed values, that is, $n_{top} = 1$ and $n_{bottom} = 2$.

In HMPEA, the precision is $2^{-m_{prec}}$, and the number of qubits needed is $n_{accur} + n_{redun} + n_{bottom}$, where $n_{bottom} = 2$. In our experiments, we adopt a resource-saving scheme in HMPEA, that is, $n_{accur} = 1$. The precision is improved by executing multiple phase estimation modules.

The red dotted line in Fig.7 indicates the maximum number (i.e. 28) of available qubits provided by PyQpanda.

### 3.1 Results of HHL

It is assumed here $n_{accur} = 9$ and $n_{redun} = 7$. Because $P$ already satisfies the normalized condition, we have $C_P = 1$. The steps are:

- Initialize the system to $|\psi_0\rangle = |0\rangle|0\rangle^{\otimes 16}|P\rangle$.
- Perform the phase estimation module, and the system state evolves to $|\psi_1\rangle = |0\rangle\sum_{j=1}^{4}p_j|\lambda_j^{[1:16]}\rangle|u_j\rangle$.
- Perform the controlled rotation, and the system state evolves to $|\psi_2\rangle = \sum_{j=1}^{4}p_j(\sqrt{1-(C_\lambda / \lambda_j^{[1:9]})^2}|0\rangle + C_\lambda / \lambda_j^{[1:9]}|1\rangle)|\lambda_j^{[1:16]}\rangle|u_j\rangle$.
- Perform the inverse quantum phase estimation, and the system state evolves to $|\psi_3\rangle = \sum_{j=1}^{4}p_j(\sqrt{1-(C_\lambda / \lambda_j^{[1:9]})^2}|0\rangle + C_\lambda / \lambda_j^{[1:9]}|1\rangle)|0\rangle^{\otimes 16}|u_j\rangle$.
- Through post-selection, the state of qubits in the bottom register evolves to $|C_\theta \cdot \tilde{\theta}\rangle = C_\lambda \sum_{j=1}^{4}p_j |u_j\rangle / \lambda_j^{[1:9]}$.

By measuring the qubits in the bottom register, the following normalized solution is obtained:

$$| C_\theta \cdot \tilde{\theta} \rangle = [0.5182 \quad 0.2843 \quad 0.3651 \quad 0.7197]^T .$$ (29)

The corresponding relative error in our numerical experiment is

$$n_t^{exp} = \frac{\left\| C_\theta \cdot \tilde{\theta} \rangle - | C_\theta \cdot \theta \rangle \right\|}{\left\| C_\theta \cdot \theta \rangle \right\|} = 0.0130 .$$ (30)

According to Eq. (7), its theoretical relative error is $n_t^{theory} = 0.0129$ .

When the number of redundant qubits $n_{redun}$ is, respectively, set to 7 and 9, and the number of accuracy qubits $n_{accur}$ is increased from 5 to 16, the relative error of the normalized solution stored in the bottom register in these 24 experiments are shown in Fig. 8(a).

As shown in Fig. 7, if $n_{redun} = 11$, the qubits needed for HHL will exceed the maximum number of available qubits provided by PyQpanda when the precision is higher than $2^{-14}$. So it can be seen from Fig. 8(a) that the experiments with $n_{redun} = 11$ are not carried out for HHL.

## 3.2 Results of HMPEA

It is assumed here that $m_{prec} = 9$, $n_{accur} = 1$, $n_{redun} = 7$, so 9 phase estimation modules are needed. Under the same precision, HHL needs 16 qubits while HMPEA needs 10 qubits. The lowest bound of the probability of successful phase estimation in HMPEA is $P_{success}^{M} = (1 - (2 \cdot (2^7 - 2))^{-1})^9 = 96.48\%$ , which is slightly lowest than that of $99.61\%$ in the HHL algorithm.

Following the procedures described in Sec. 2.2, $\lambda_j^{[1:9]}$ , $C_P^2 p_j^2$ and the absolute value of each element of $u_j$ are extracted and shown in Table 2.

**Table 2**
Experimental results of HMPEA.

| Phase Estimation - index | | $\lambda_1$ | $\lambda_2$ | $\lambda_3$ | $\lambda_4$ |
|---|---|---|---|---|---|
| Phase Estimation - 1 | | $\lambda_1^{[1:3]} = 101$ | $\lambda_2^{[1:3]} = 011$ | $\lambda_{\{3,4\}}^{[1:3]} = 000$ | |
| Phase Estimation - 2 | | $\lambda_1^{[4:6]} = 110$ | $\lambda_2^{[4:6]} = 011$ | $\lambda_3^{[4:6]} = 111$ | $\lambda_4^{[4:6]} = 010$ |
| Phase Estimation - 3 | | $\lambda_1^{[7:9]} = 000$ | $\lambda_2^{[7:9]} = 011$ | $\lambda_3^{[7:9]} = 011$ | $\lambda_4^{[7:9]} = 110$ |
| Joint statistics | Eigenvalues | $\lambda_1^{[1:9]} = 101110000$ | $\lambda_2^{[1:9]} = 011011011$ | $\lambda_3^{[1:9]} = 000111011$ | $\lambda_4^{[1:9]} = 000010110$ |
| | Joint probability | $C_P p_1^2 = 0.3681$ | $C_P p_2^2 = 0.3001$ | $C_P p_3^2 = 0.2114$ | $C_P p_4^2 = 0.0921$ |
| Eigenvectors | | $abs(u_1) = \begin{bmatrix} 0.7444 \\ 0.1296 \\ 0.0497 \\ 0.6531 \end{bmatrix}$ | $abs(u_2) = \begin{bmatrix} 0.0298 \\ 0.6986 \\ 0.6973 \\ 0.1579 \end{bmatrix}$ | $abs(u_3) = \begin{bmatrix} 0.5356 \\ 0.3226 \\ 0.4458 \\ 0.6406 \end{bmatrix}$ | $abs(u_4) = \begin{bmatrix} 0.3976 \\ 0.6253 \\ 0.5593 \\ 0.3716 \end{bmatrix}$ |

The approximate solution of the DC power flow equations is obtained by Eq. (19) as

$$\tilde{\theta} = [0.0084 \quad 0.0046 \quad 0.0059 \quad 0.0116]^T .$$ (31)

The corresponding relative error is

$$n_t^{exp} = \frac{\left\| \tilde{\theta} - \theta \right\|}{\left\| \theta \right\|} = 0.0242 .$$ (32)

Due to the failure possibility of phase estimation, it deviates slightly from the theoretical relative error $n_t^{theory} = 0.0285$ . In HMPEA, totally 36 experiments are performed, in which $m_{prec}$ increases from 5 to 16 with $n_{accur} = 1$ and $n_{redun}$ equal to 7, 9 or 11. The relative error is shown in Fig. 8(b).

HHL improves the precision by using more accuracy qubits while HMPEA by executing more phase estimation modules. Therefore, the experiments of HMPEA with $n_{redun} = 11$ can be carry out in Fig. 8(b), which shows the superiority of HMPEA in saving qubit resources when achieving comparable precision.

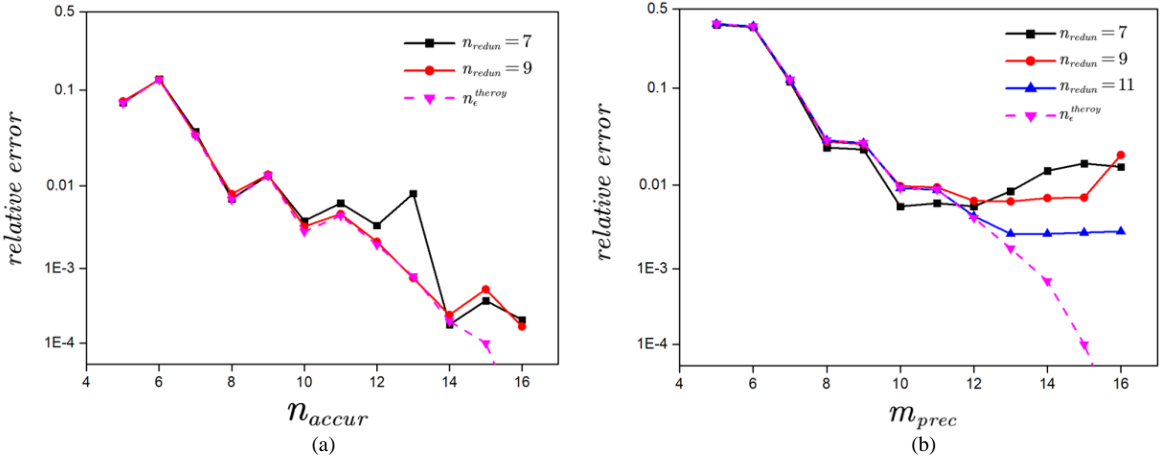

(a)                                                                 (b)

**Fig. 8.** Relative error curves of (a)HHL and (b)HMPEA

### 3.3 Discussions

In the problem of solving the linear system of equations, the time complexity of the best classical algorithm is $O(Nsk\log(1/\epsilon))$ , where $N$ is the dimension of the system, $s$ is the sparsity of the matrix, $k$ is the condition number, $\epsilon$ is the theoretical error we can tolerate. HHL can achieve exponential speedup with the time complexity $O(\log(N)s^2k^2/\epsilon)$ . Since HMPEA employs $\lceil m_{prec}/n_{accur}\rceil$ phase estimation modules, the time complexity for HMPEA should be roughly $O(\log(N)\lceil m_{prec}/n_{accur}\rceil s^2k^2/\epsilon)$ .

For HHL under imperfect phase estimation, a large number of qubits is required to get an accurate solution. HSPEA solves the linear system of equations by combining the phase estimation and measurements, which can reduce the circuit depth with roughly the same number of qubits. HMPEA reduces the demand for the number of accuracy qubits by using multiple phase estimation modules and the circuit depth of each module is the same as that of HSPEA. Compared with HHL, HMPEA transfers the complexity to the measurement, and an efficient measurement scheme will improve the performance of HMPEA.

When the HHL algorithm is used to solve DC power flow equations, the error mainly comes from the failure of phase estimation induced by insufficient number of redundant qubits and the truncation of eigenvalues induced by insufficient number of accuracy qubits. When the number of accuracy qubits is small, the deviation caused by eigenvalue truncation is the dominant factor of the error, and the error of the solution can be reduced by increasing the accuracy qubits under the condition that the number of redundant qubits is constant. When the number of accuracy qubits increases to a certain extent, the failure of phase estimation becomes the main error source, and the error can be reduced by adding redundant qubits. As shown in Fig. 8(a), when there are enough redundant qubits, the relative error curve in experiment fits well with that in theory. In general, when using the HHL algorithm to solve the DC power flow problem, the error of the solution can be reduced by increasing the number of accuracy and redundant qubits in the medium register, that is, the accuracy can be improved with more qubit resources. Besides, we can see that to achieve comparable precision with the HHL algorithm, HMPEA can solve DC power flow equations with less qubit resources, which indicates that HMPEA may be a feasible route for solving linear systems of equations under imperfect phase estimation in the NISQ era.

The relative error curves of HHL and HMPEA are intuitive and demonstrate that more qubits or more phase estimation modules can achieve a higher solution accuracy. For HHL, the increase of the number of accuracy and redundant qubits can make the relative error close to 0. While for HMPEA, the increase of the estimation precision of the eigenvalues (i.e. the number of accuracy qubits multiply that of phase estimation modules) and the number of redundant qubits can make the relative error close to 0. When the precision is higher than $2^{-13}$ (i.e. $m_{prec} = 13$), the error mainly comes from the failure rate of phase estimation, so the experimental relative errors of HMPEA gradually failed to fit with the theoretical relative error.

## 4. Conclusion

In this paper, the HMPEA algorithm is developed to solve DC power flow problems with fewer qubits. The IEEE 5-bus test system is taken as an example to illustrate the performance of the HHL and HMPEA algorithms in DC power flow problems. For HHL, the relationship between the relative error of the solution and the qubit resources (i.e. the number of accuracy and redundant qubits) is discussed. The demand for the number of accuracy qubits is reduced in HMPEA by using multiple phase estimation modules. The numerical experiments show that HMPEA can achieve comparable accuracy as HHL with fewer qubits when solving DC power flow problems, which may provide a feasible route in the NISQ era.

### Acknowledgments

This work was supported by the National Natural Science Foundation of China [grant numbers 61720106009 and 61773359].